\documentclass[journal=mamobx,manuscript=article]{achemso}

\usepackage{multirow}
\usepackage{booktabs}
\usepackage{color,soul}
\usepackage{lipsum}
\usepackage{tikz}
\usepackage{graphicx}
\usepackage{placeins}
\usepackage[version=3]{mhchem} 

\author{Utku G\"urel}
\author{Andrea Giuntoli}
\affiliation{University of Groningen, Zernike Institute for Advanced Materials, Nijenborgh 4, 9747AG Groningen, The Netherlands}
\email{a.giuntoli@rug.nl}

\title[An \textsf{achemso} demo]
  {Shear Thinning from Bond Orientation in Model Unentangled Bottlebrush Polymer Melts}

\begin{document}

\newpage
\section{TOC Graphic}
\begin{figure}
\includegraphics{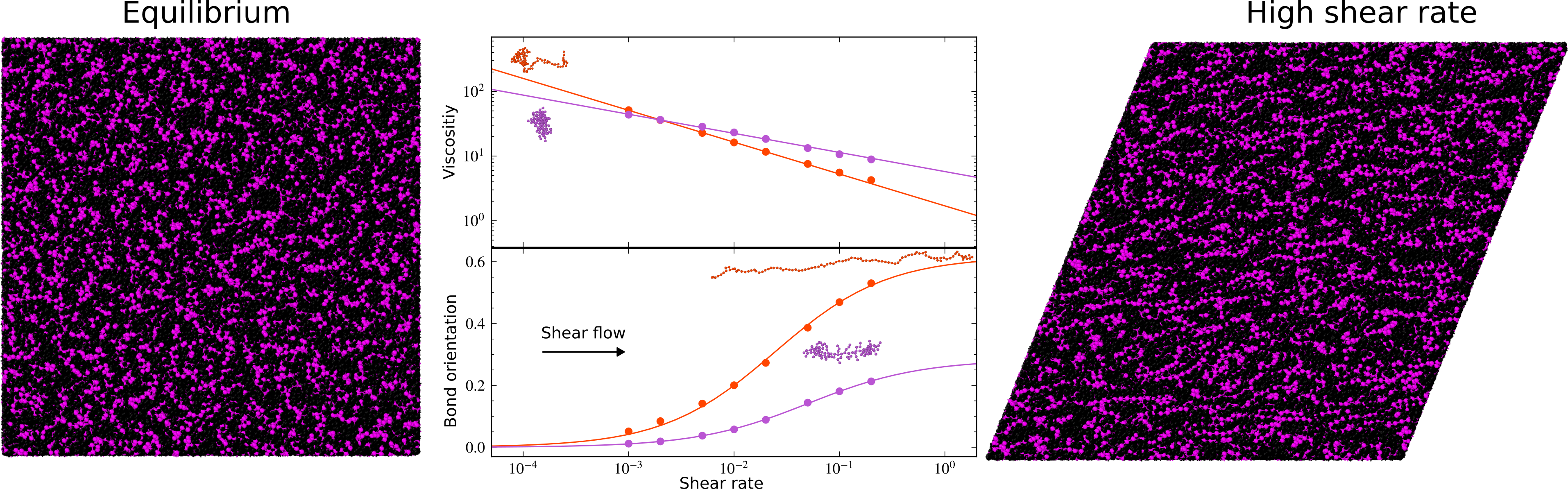}
\end{figure}


\newpage
\begin{abstract}
The rheology of molecular brushes remains challenging to control due to the multiple length scales and relaxation processes involved, and the lack of direct observation of molecular conformation during flow. We use molecular dynamics simulations to determine the shear thinning of unentangled bottlebrush polymers with varying architecture, from linear chains to combs, to densely grafted bottlebrushes, to star-like and star polymers. We find shear thinning exponents in line with theoretical and experimental results and characterise the shape and orientation of bottlebrushes in steady-state flow. Many shape parameters derived from the gyration tensor show molecular alignment with the flow for all systems. Yet, the orientation of individual bonds is what most strongly correlates with the architecture-dependent shear-thinning exponents. In densely grafted bottlebrushes, the packing of side chains prevents alignment with the flow causing a reduction in shear thinning. The molecular insight from our simulations is useful to tune the architecture of bottlebrushes to control their rheology.
\end{abstract}
\newpage
\section{Introduction}

Bottlebrushes are a type of branched polymer composed of a linear backbone and multiple side chains grafted to it. Their characteristic architecture and the dense packing of side chains lead to unique properties that have been studied over the last few decades \cite{dziezok1997cylindrical, saariaho1998extended, subbotin2000elasticity, lecommandoux2002effect, sheiko2008cylindrical, hsu2010characteristic, verduzco2015structure, daniel2016solvent, bichler2021dynamical} . The shape and the size of bottlebrushes are determined by three molecular parameters: the length of the backbone,  the length of the side chains, and the spacing between these side chains (grafting density). At equilibrium, a bottlebrush is more rigid and extended than a linear chain with the same backbone length due to the side chains' steric repulsion, leading to higher persistence length, stiffness, and increased packing efficiency \cite{borodinov2018molecular} .
 The overall shape of a bottlebrush polymer is highly dependent on the interplay of architectural parameters. Increasing grafting density results in more compact and cylindrical molecular shapes. For a fixed grafting density, increasing side-chain length stretches the backbone chain and the bottlebrush becomes more rigid \cite{sheiko2008cylindrical} . 
This rich behaviour leads to highly tunable mechanical and thermal properties of bottlebrushes compared to linear polymer chains \cite{paturej2016molecular} and to softer and more flexible networks \cite{cai2015soft} . The control over the architectural composition makes bottlebrushes suitable candidates for multiple applications in biomedical sciences  \cite{walsh2019engineering} , soft robotics \cite{sarapas2018compressing}  , pressure-sensitive adhesives \cite{arrington2018reversibly} , capacitive pressure sensors \cite{reynolds2020super} , and ultrasoft electronics \cite{xu2023conductive} . 

While the configuration of bottlebrushes at equilibrium is somewhat understood \cite{chremos2018comparative, paturej2016molecular} , it remains challenging to predict their shape rearrangement under mechanical deformations and flow. This prediction is necessary to determine their rheological behaviour and to use them in technological applications such as 3D printing \cite{patel2020tunable} . It is well-known, for example, that polymer melts exhibit shear thinning \cite{ferry1980viscoelastic} due to the alignment of the polymer chains with the flow direction \cite{forster2005shear} . In branched polymers, the shear thinning is governed by multiple length scales and relaxation times \cite{doi1988theory,fetters1993rheological} , which gives rise to unique rheology compared to linear chains \cite{ripoll2006star, nikoubashman2010branched, abbasi2017linear, lopez2019extensional, wever2013branched} that is not trivial to predict from the molecular architecture. This is complicated by the challenges in the controlled synthesis and characterisation of branched polymers \cite{haugan2018consequences, van2014molecular} . 

Several studies have been conducted to investigate the rheological properties of melts or solutions of bottlebrushes (synthetic or biological) such as their viscoelasticity \cite{yavitt2017high, golkaram2020supramolecular} , shear-induced crystallinity \cite{nagaraj2021strain, clarke2022bottlebrush} , and shear-thinning behaviour in aggrecan solutions \cite{horkay2021rheological} . A few experimental studies shed light on the complex contribution of the bottlebrush architectural parameters on their melt rheology. Hu \textit{et al.} showed first the relative importance of the different relaxations of side chains and backbones \cite{hu2011linear} . Dalsin \textit{et al. }reported an increase in zero-shear viscosity for bottlebrushes of increasing backbone length and fixed side chain length \cite{dalsin2014molecular} , with weak molecular weight scaling at the low molecular weight (corresponding to a star-like regime, where the backbone and the side chains are of comparable length), and a Rouse-like scaling for longer backbones, similar to linear unentangled chains. The same group identifies a star-like to bottlebrush-like transition and recently reported that this transition also signals the onset of a strain-hardening regime in extensional flow \cite{zografos2023star} . L\'{o}pez-Barr\'{o}n \textit{et al.} found that the complex viscosity of unentangled bottlebrushes decreases with increasing side-chain length, due to the decrease of backbone-backbone friction \cite{lopez2015linear} . They recently attributed a similar strain-hardening of bottlebrushes in extensional flow to the entanglement of side chains leading to side chain interdigitation and alignment perpendicular to the flow \cite{lopez2019extensional} , though Zografos \textit{et al}. show that this effect is present even when side chains are not entangled \cite{zografos2023star} .
Similar scalings for the linear viscoelasticity of star-like bottlebrushes have also been reported by Alexandris \textit{et al.} \cite{alexandris2020linear} , linking the rheological behaviour to the expected relaxation times of bottlebrushes based on their expected equilibrium configuration.  

Despite the numerous observations and hypothesised behaviour, it remains challenging to experimentally observe the molecular orientation with the melt flow of individual bottlebrushes (and branched polymers in general). Computational studies become then invaluable to precisely determine the architectural and shape effects of branched polymers in the flow properties of melts. A few computational papers studied the conformation of nonlinear polymers such as rings \cite{xu2015simulation, o2020topological} , stars \cite{xu2015simulation, ripoll2006star} , or comb-like squalene molecules \cite{kadupitiya2021probing} . Qu \textit{et al.} briefly report a decreasing alignment with the flow direction related to a decrease in shear thinning for bottlebrushes with increasing backbone length, but to our knowledge, there are no computational studies of the effect of chain topology on the conformation under flow and shear thinning of unentangled bottlebrush polymer melts. 

The primary goal of this work is to study the effect of bottlebrush architecture on the viscosity of the melt for unentangled chains under shear flow. We use non-equilibrium molecular dynamics (NEMD) simulations to calculate the flow curve in steady shear for different bottlebrush architectures at a fixed molecular weight and compare them with the limit cases of linear chains, comb polymers, and stars. We demonstrate that all polymer melts exhibit shear thinning, and we also observe a transition from star-like to bottlebrush-like structures. However, in the investigated frequency range, we find that the thinning exponent is determined by the orientation with the flow of individual bonds rather than chain orientation. Densely grafted bottlebrushes shear-thin the least because the packing of the side chains prevents their strong alignment with the flow direction, and this effect decreases with decreasing grafting density from the bottlebrush to the comb regime. Many other shape parameters that are derived from the principal components of the gyration tensor also show molecular alignment with the flow but do not directly correlate with the shear thinning behaviour as a function of molecular architecture. Our simulations provide important insight into the bond and chain configuration changes under shear flow that are crucial to predict and control the rheology of bottlebrush polymer melts in different architectural regimes.

\section{Methods}
In this work, we use coarse-grained molecular dynamics (CGMD) simulations under non-equilibrium conditions.

\subsection{Model}
We employ the bead-spring model with reduced Lennard-Jones (LJ) units where the energy ($\varepsilon$), mass ($m$), and diameter ($\sigma$) of a single bead are set to $1$ leading to a unity time scale as $\tau = \sqrt{m\sigma^2 / \varepsilon}$ \cite{kremer1990dynamics} . The bonded interactions in the bead-spring model are given by the finite extensible nonlinear elastic (FENE) potential

\begin{equation}
    U_{fene}(r) = -0.5k{R_0}^2 ln \left[ 1- \left(\frac{r}{R_0}\right)^2 \right ]
\end{equation}

\noindent to represent the bonds between connected monomers, where $k=30\varepsilon/\sigma^2$ is the spring constant, $R_0=1.5\sigma$ is the maximum allowed bond length, and $r$ is the distance between two bonded beads. Non-bonded interactions between the beads are given by the truncated LJ potential

\begin{equation}
    U_{LJ}(r)=\left\{\begin{array}{lc}
\varepsilon\left[\left(\frac{\sigma^{\text{*}}}{r}\right)^{12}-\left(\frac{\sigma^{\text{*}}}{r}\right)^6\right], & r \leq r_{c} \\
0, & r>r_{c}
\end{array}\right.
\end{equation}

\noindent to represent the Van der Walls interactions up to a cutoff distance $r_c=2.5\sigma$, where $\sigma^{\text{*}}=2^{1/6} \sigma$ is the length scale at which the LJ potential attains its minimum value with depth $\varepsilon$. CGMD simulations are carried out with the LAMMPS (Large-scale Atomic/Molecular Massively Parallel Simulator) software with periodic boundary conditions in all three dimensions. (https://www.lammps.org/)\cite{thompson2022lammps} . Polymer chain snapshots are rendered with OVITO \cite{stukowski2009visualization} .

\subsection{Chain Parameters}
We analyse bottlebrush architectures with varying backbone ($N_{BB}$) and side chain ($N_{SC}$) lengths, specifically, $ \{(N_{BB}, N_{SC}, m)\} = \{((25, 3, 1), (20, 4, 1), (10, 9, 1), (5, 19, 1) \}  $, leading to a total number of beads $N_{BB} (N_{SC}+1) = 100 $ in a single chain with $m=1$, $i.e$ one side chain per backbone bead. The total number of chains is $2,000$ making a total of $200,000$ beads in the melt. Bottlebrushes with grafting density $m>1$ have the following parameters: $ \{(N_{BB}, N_{SC}, m)\} = \{(49, 4, 4), (33, 4, 2) \}  $. Here, we graft a side-chain to every $m^{th}$ backbone bead. Both of these systems have a molecular weight of $M_w=101$, and there are $1,980$ chains making the total number of beads $199,980$ as close as possible to the other systems. Additionally, we compare bottlebrushes to a system of linear chains with length $L=100$ and a system of star polymers with $f=9$ arms and $M=11$ beads per arm. Note that these two additional architectures also satisfy the constraint we set on the molecular weight $M_w=100$ with $2,000$ chains. The chain parameters are summarised in Table~\ref{tab:architectures}.

\begin{table}[ht]
\small
  \caption{\ Architectural parameters for bottlebrushes with backbone length $N_{BB}$ and side chain length $N_{SC}$, linear polymers of length $L$, star polymers of $f$ arms with arm length $M$. The molecular weight is given by $N_{BB} (1+N_{SC}/m)$ for bottlebrushes and by $fM+1$ for stars.}
  \label{tab:architectures}
\begin{tabular}{@{}ccccc@{}}
\toprule
Topology                     & $N_{BB}$   & $N_{SC}$  & $m$ \\ \midrule  
\multirow{8}{*}{Bottlebrush} & 25         & 3         & 1 \\     
                             & 20         & 4         & 1  \\   
                             & 10         & 9         & 1  \\   
                             & 5          & 19        & 1  \\   
                             & 33         & 4         & 2  \\  
                             & 49         & 4         & 4  \\   
Linear                       & \multicolumn{3}{c}{L=100} \\ \midrule  
Star                         & \multicolumn{3}{c}{f=9,  M=11} \\ \bottomrule 
\end{tabular}
\end{table}

\subsection{Equilibration protocol}

To prepare the polymer melt, we first randomly pack the desired number of polymer chains in the simulation box followed by an energy minimisation step. Then we equilibrate the system in the NPT ensemble (constant number of particles, pressure, and temperature) using a Nose-Hoover thermostat at temperature $<T>=1.0$ and pressure $<P>=0.1$. We choose this pressure value since it is close to atmospheric pressure \cite{liu2021effects} . We run an initial equilibration for $5\times 10^7$ simulation steps and additional $5\times10^6$ production steps with an integration step of $0.005\tau$ ensuring that the initial end-to-end vector correlations of polymer chains are lost (see Fig. S1). We consider the longer subchains of bottlebrushes (either the backbone or the side chain) in the correlation calculations. For star polymers, we consider the chain segment from the core to a terminal arm bead. The correlations are averaged over all subchains and for all chains in the melt. We also have a final average over 3 independent simulation runs each of which starts with a different initial configuration of polymer chains at different random initial velocities.

\subsection{Deformation protocol} 

After the creation of the polymer melt, we apply a simple shear with a fixed shear rate on the $xy-$plane where the directions of shear, gradient, and vorticity are $\hat{x}$, $\hat{y}$, and $\hat{z}$ respectively. The SLLOD equations of motion are integrated with Lees-Edwards boundary conditions at constant volume. We measure the average stress values $\sigma_{xy}$ of all beads in the system which is a function of the total strain $\gamma = \dot{\gamma}t$. We then calculate the viscosity $\eta=\sigma_{xy} / \dot{\gamma}$ of each system after deforming them up to a total strain of $\gamma=100$. The overall shearing procedure is the same as in Ref.~\cite{giuntoli2020predictive} . We explore a range of shear rates from $\dot{\gamma} = 10^{-3}$ to $\dot{\gamma} = 0.2$ similar to Ref.~\cite{peng2021rheological} observing unstable viscosity data for shear rates lower than $10^{-3}$ as they also report in their work. We observe an unexpected drop in the pressure of the system when we try to simulate at lower shear rates. Also, at shear rates larger than $0.2$, the bonds in the system become unrealistically large and cause the simulations to fail. To ensure the stability of our simulations at high shear rates, we reduce the integration time step to $0.001\tau$ and check that the temperature fluctuates stably around the initially set value.

\subsection{Definition of architectural parameters}

We computed the following architectural parameters, most of which are a function of gyration tensor eigenvalues. The correlation between these parameters is shown in Fig. \ref{fig:spearman}. We plot them as a function of the shear rate in Fig. S3 along with the corresponding eigenvalues in Fig. S4.

\begin{itemize}
    \item Orientation resistance $\tan 2\phi = \frac{<G_{xy}>}{<G_{xx}>-<G_{yy}>}$
    \item Asphericity $b = \lambda^2_{max} - \frac{1}{2} \left ( \lambda^2_{min} + \lambda^2_{mid} \right )$
    \item Acylindricity $c = \lambda^2_{mid} - \lambda^2_{min} $
    \item Prolateness $p = \frac{\left(2\lambda_{min}-\lambda_{mid}-\lambda_{max}\right ) \left (2\lambda_{mid} -\lambda_{min}-\lambda_{max}\right ) \left ( 2\lambda_{max}-\lambda_{min}-\lambda_{mid}\right ) } { 2 \left ( R^2_g -\lambda_{min}\lambda_{mid}-\lambda_{mid}\lambda_{max}-\lambda_{min}\lambda_{max}  \right )    }  $
    \item Shape anisotropy $\kappa =  \frac{b^2 + \frac{3}{4} c^2}{R^4_g} $
\end{itemize}

\section{Results and Discussion}

\begin{figure}[h!]
\includegraphics[width=0.75\textwidth]{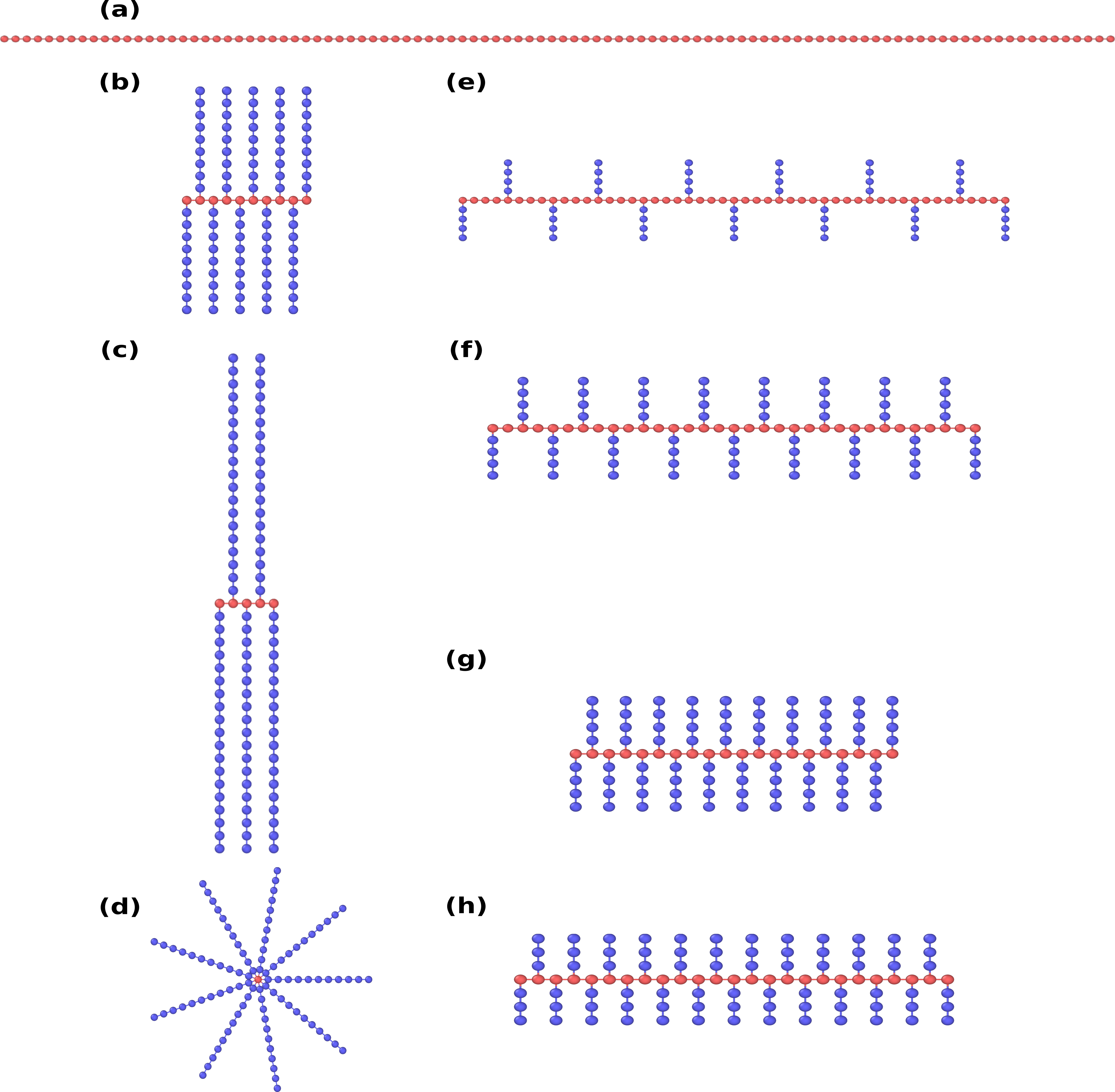}
  \caption{Simulated polymer architectures at fixed molecular weight $M_w=100$. (a) Linear chain of length 100. (b) Sphere-like polymer with $N_{BB}=10, N_{SC}=9, m=1$. (c) Star-like polymer with $N_{BB}=5, N_{SC}=19, m=1$. (d) Star polymer with $f=9, M=11$. (e) Bottlebrush with $N_{BB}=49, N_{SC}=4, m=4$. (f) Bottlebrush with $N_{BB}=33, N_{SC}=4, m=2$. (g) Bottlebrush with $N_{BB}=20, N_{SC}=4, m=1$. (h) Bottlebrush with $N_{BB}=25, N_{SC}=3, m=1$.  We focus on capturing the linear-to-star transition by increasing $N_{SC}/N_{BB}$ ratio (left column) and the linear-to-bottlebrush transition (right column) by decreasing m.} 
  \label{fig:architectures}
\end{figure}

Figure \ref{fig:architectures} shows the architectures studied. We change chain branching to obtain star, comb and bottlebrush polymers to study the topology effect. To avoid molecular weight contributions, all architectures have the same number of beads ($M_w =100\pm 1$), which also constrains our choice for the limit architectures of linear chains and star polymers. Bottlebrush and comb architectures are defined by three parameters: (i) the number of backbone beads $N_{BB}$, (ii) the number of side chain beads $N_{SC}$, and (iii) the spacing between grafted side chains ($m$). Hence, the set $ \{(N_{BB}, N_{SC}, m)\} $ encodes all architectural information. Densely grafted bottlebrush regime corresponds to $m=1$ and can be classified into three regions depending on the ratio $N_{BB}/N_{SC}$ : Bottlebrush-like, sphere-like, and star-like \cite{chremos2018comparative} . Bottlebrush-like polymers have significantly larger backbone lengths compared to their side chains ($N_{BB} >> N_{SC}$) which allow us to consider the backbone as a diluted linear chain. Sphere-like polymers have comparable backbone and side chain lengths ($N_{BB} \sim N_{SC}$) and the backbones of star-like polymers are much shorter than their side chains ($N_{BB}<N_{SC}$). The architectures with $m \geq N_{SC}$ correspond to loosely grafted combs, and $m \approx N_{SC}$ marks the transition region between loosely grafted combs to densely grafted combs \cite{paturej2016molecular} .  

\subsection{Shear thinning of polymer melts}

\begin{figure}[h!]
  \includegraphics{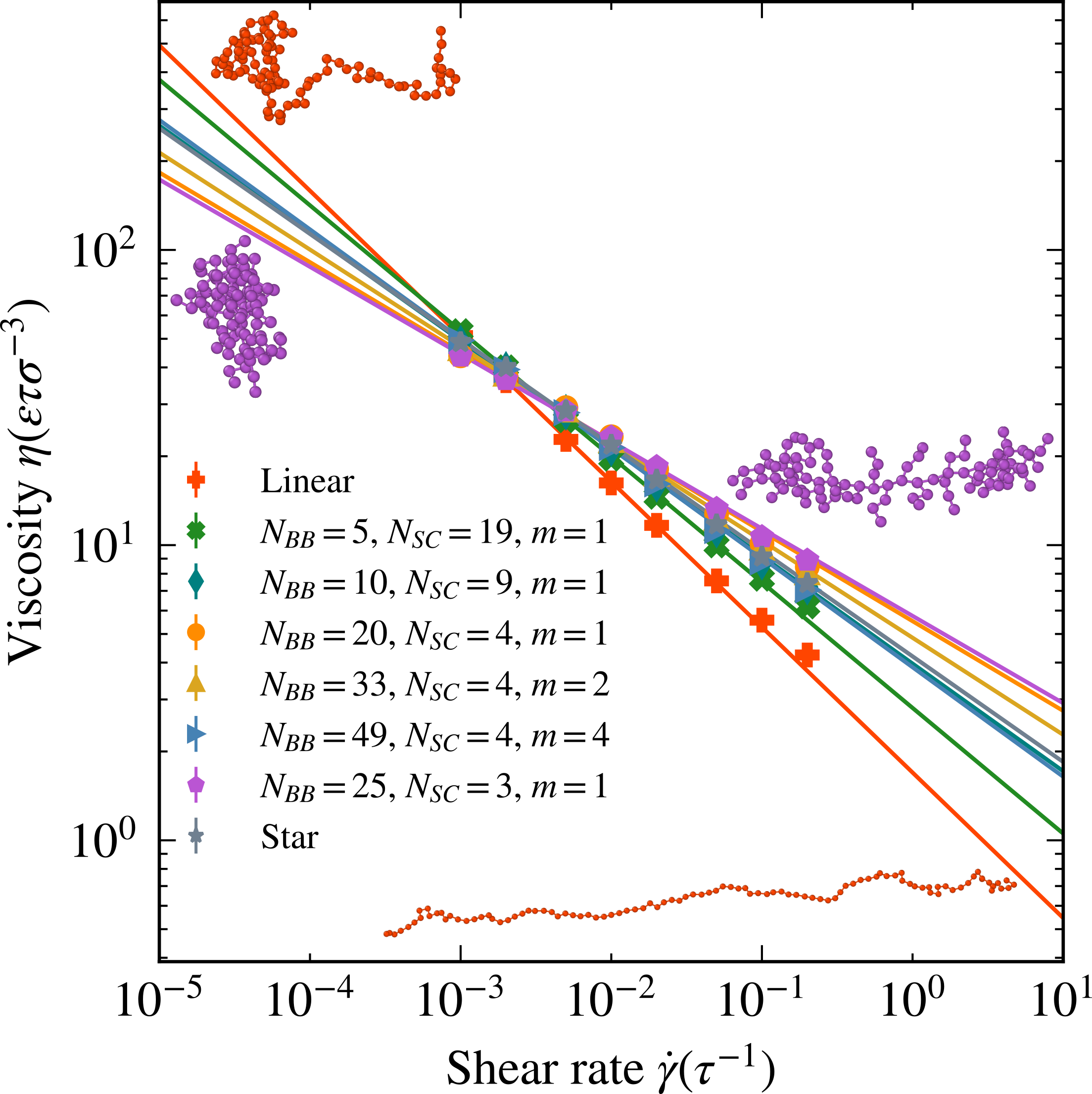}
  \caption{Shear viscosity of different architectures as a function of shear rate. Data are shown by solid markers, and the lines are exponential fit to the data given by the expression $\eta = K \dot{\gamma}^{n-1}$. The drop in viscosity with increasing shear rate indicates shear thinning for all systems. We observe an architectural difference between systems with the same molecular weight, with the strongest and weakest shear thinning for linear chains and dense bottlebrushes, respectively. The snapshots show a typical configuration of a single bottlebrush chain of $N_{BB}=25$, $N_{SC}=3, m=1$ and a linear chain at the lowest (left) and the highest (right) shear rates. The simulated architectures are elongated  and aligned with the flow at high shear rates, and they have a more globular shape at low shear rates.}
  \label{fig:viscosity}
\end{figure}

Here we show the shear thinning behaviour of polymer melts with different molecular architectures. We calculate the viscosity as 

\begin{equation}
    \eta(\dot{\gamma}) = \frac{\sigma_{xy}}{\dot{\gamma}},
\end{equation}

\noindent where $\sigma_{xy}$ is the shear stress at steady flow given by the average per-monomer stress, and $\dot{\gamma}$ is the shear rate. The viscosity flow curves of polymer melts with different architectures are reported in Fig. \ref{fig:viscosity} and show a clear shear thinning trend. We fit the data with a power-law model $\eta = K \dot{\gamma}^{n-1}$ where $n$ is the shear thinning exponent and $K$ is the flow consistency index. Shear thinning exponents for all systems are reported in Table \ref{tab:exponents}. The value of $n$ ranges between 0 and 1 and smaller values of $n$ indicate stronger shear thinning. The prediction of this exponent for unentangled linear chains is $n=0.5$ which is known from the theory of polymer physics and confirmed by experiments \cite{rubinstein2003polymer, colby2007shear} . We can confirm that the linear chains in our system are not entangled by the measurement of the scaling exponent $n$. Since the bottlebrushes in our systems are unentangled, we also choose unentangled linear chains to make a better comparison between different architectures. Note that for entangled systems, the initial drop in the viscosity during the start-up shear is attributed to the chain disentanglement \cite{wang2006nonquiescent} . Hence, we exclude this mechanism by choosing unentangled linear chains, as later proven by their shear-thinning behaviour (see Table \ref{tab:exponents} ). We would expect a kink in the viscosity curve around $\dot{\gamma} \sim 10^{-4}$ based on the relaxation times $\tau_R$ of the longest subchain of our polymers (see Fig. S1). For shear rates lower than $\tau^{-1}_R$, we would reach the Newtonian plateau. Although we try to reach this plateau, the viscosity data fluctuate largely for $\dot{\gamma} < 10^{-3}$, and the simulations become unstable. Nevertheless, based on the relaxation times $\tau_R$ in Fig. S1 and Rouse model arguments for unentangled polymers, we might expect a higher viscosity at lower shear rates for linear chains and other elongated architectures, given that the total molecular weight is fixed \cite{rubinstein2003polymer} . A crossover point is then expected at intermediate shear rates around $\dot{\gamma} \sim 1/\tau_R$, which is roughly what we observe in Fig. 2.

One would expect stronger shear thinning for architectures that better align with the flow at an increasing shear rate. Yet, for complex branched architectures the interpretation of what "alignment with the flow" means is not trivial. In the following sections we thoroughly analyse the dependence on the shear rate of multiple shape parameters and their relation with the melt shear thinning behaviour, but we can already make some observations based on the viscosity data.
Linear chains easily elongate in the flow direction and show the strongest shear thinning. Densely grafted architectures (a-b-c-d-g-h in Fig. \ref{fig:architectures}) do not show a clear trend with varying $N_{BB}/N_{SC}$ ratio from the limits of linear chains to star polymers. Instead, bottlebrush-like systems (architectures g and h) show the weakest shear thinning behaviour.
Another observation we can make is the effect of grafting density on shear thinning. Comparing the viscosity data in Fig. \ref{fig:viscosity} for the topologies with the same side-chain length ($N_{SC}=4$), the loosely grafted architecture with $m=4$ has the lowest viscosity at the high shear rate, followed by more densely-grafted architectures with $m=2$ and $m=1$ respectively. Therefore, the increasing spacing between grafted side chains enhances the shear thinning. We discuss the molecular origin of these trends in the following sections.

\begin{table}[h!]
\caption{Shear thinning exponent \textit{n} for different architectures}
\label{tab:exponents}
\begin{tabular}{ccc}
\toprule
Architecture            & $n$       \\
\midrule
$Linear$                 & 0.507     \\
$N_{BB}=5$, $N_{SC}=19, m=1$ & 0.574 \\
$N_{BB}=49$, $N_{SC}=4, m=4$ & 0.629 \\
$N_{BB}=10$, $N_{SC}=9, m=1$ & 0.635 \\
$Star$                       & 0.642 \\
$N_{BB}=33$, $N_{SC}=4, m=2$ & 0.671 \\
$N_{BB}=20$, $N_{SC}=4, m=1$ & 0.696 \\
$N_{BB}=25$, $N_{SC}=3, m=1$ & 0.704 \\

\bottomrule
\end{tabular}
\end{table}

\subsection{Bond orientation parameter}

Orientational order is the driving mechanism in the shear thinning of complex fluids \cite{forster2005shear} , but its definition for branched polymers is not trivial. In fact, one would expect the shear thinning to be correlated to some metric derived from the gyration tensor, but we report later in detail that this is not the case for our system. Instead, we find a strong correlation between shear thinning and the orientation of the bonds. To quantify it, we measure the bond orientation parameter $<P_2>$ which is defined as   

\begin{equation}
    <P_2(\cos \theta)> =  \frac{3\left<\cos^2\theta \right > -1 }{2} 
\end{equation}

\noindent where $P_2$ is the second order Legendre polynomial, $\theta$ is the angle between the bond vector and the unit vector $\hat{x}$. The average $<.>$ is taken over all bonds in a single chain and over all chains in the system. We also calculate $P_2$ for backbones and side chains separately (see Fig. S2). $<P_2>$ takes values between $-0.5$ and $1$ where $<P_2>=-0.5$ indicates a perpendicular alignment to the flow direction, and $<P_2>=1$ a perfect parallel alignment. Under equilibrium conditions,  $<P_2>$ becomes $0$ indicating a random orientation without any directional preference. We also measure and report $<P_2>$ at equilibrium in Table \ref{tab:equilShape} confirming that the orientation under equilibrium conditions is $0$ as expected.

\begin{figure}[h!]
  \includegraphics{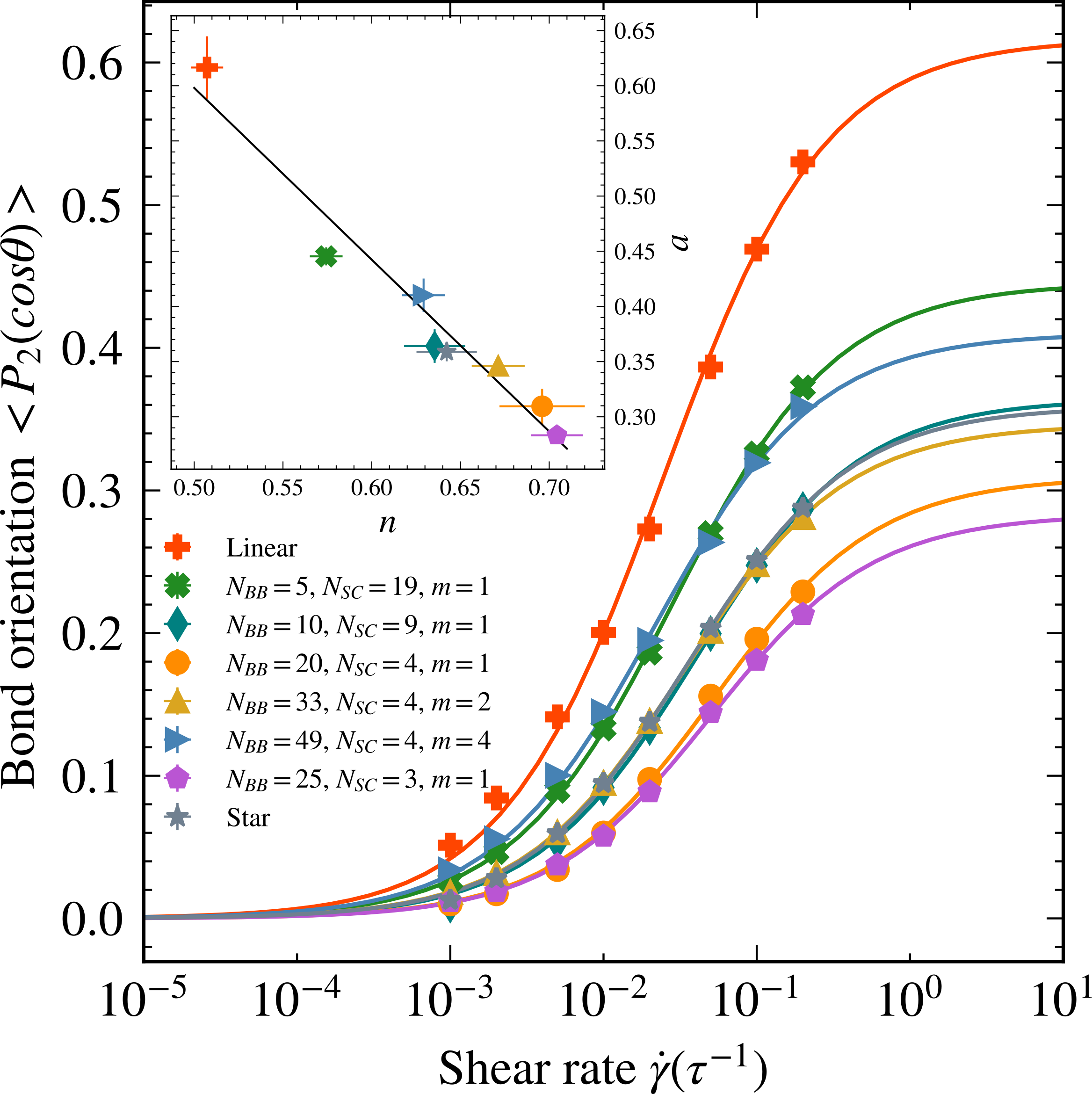}
  \caption{Bond orientation parameter of different architectures as a function of shear rate. Data are shown by solid markers and the lines are generalised logistic function fits. Individual bonds in polymer chains align better with the flow at high shear rates, leading to an increase in shear thinning. Inset: Predicted value $a$ of bond orientation at high shear rate regime as a function of shear thinning exponent $n$. Fit parameters are shown with solid markers, and the error in the fit is given by twice the standard deviation. Inverse linear relation indicates that the shear thinning increases when the bonds orient more with the flow.}
  \label{fig:orientation}
\end{figure}

We show the values of $<P_2>$ at all shear rates in Fig. \ref{fig:orientation} with solid markers. $<P_2>$ converges to $0$ at low shear rates, recovering the equilibrium value, whereas it increases with increasing shear rate as a result of an increased bond orientation. We expect $P_2$ to reach a plateau at the infinite shear rate limit imposed by both topological constraints and the chain stiffness \cite{lang2019microstructural} . Although the perfect alignment with flow indicates a value of 1 for $P_2$, this limit is never reached because of the entropic contributions on the order of $\sim k_B T$. The interactions between different chains prevent them from moving on a straight line on the x-axis and hence aligning perfectly. Plateau values of $P_2$ have been previously reported up to $0.5$ for squalane, a comb-like unentangled polymer, at different thermodynamic conditions \cite{jadhao2019rheological} and up to $0.8$ for rigid colloidal rods \cite{lang2019microstructural, ripoll2008attractive} . Note that reaching this plateau experimentally is not possible due to the high shear rates required, and even our simulations become numerically unstable beyond the data points presented here. 
Nevertheless, we can fit our data with the generalised logistic function (solid lines) that is given by the expression $P_2(\dot{\gamma}) = a + \frac{-a}{1+ \left ( \frac{\dot{\gamma}}{c} \right )^{0.82} }$. The fit parameter $a$ describes the behaviour at the infinite shear rate regime when the bond orientation reaches a plateau, and $c$ is a characteristic shear rate at which bonds transition from their equilibrium configuration to an alignment with the flow. $1/c$ is roughly related to the inverse of the bond relaxation time, on the order of $~10^2\tau$, as we discuss in Fig. S1(c). The exponent of $\frac{\dot{\gamma}}{c}$ determines how sharp the transition is from $0$ to $a$. The value $0.82$ is the best fit to data and is fixed for all systems. The fact that a single exponent is sufficient for all systems indicates that it does not strongly depend on the architecture, though we don't have a particular insight into this specific parameter.

In the range of shear rates studied, the progressive bond orientation with the flow governs the shear-thinning behaviour of all architectures: the higher the bond orientation, the stronger the shear thinning, see the inset of Fig. \ref{fig:orientation}. We observe that bottlebrush orientation increases with increasing $m$ for the same side-chain length ($N_{SC}=4$). The loosely-grafted bottlebrushes with $m=4$  orient better than more densely-grafted bottlebrushes with $m=2$ and $m=1$. This trend is the same as their high shear rate viscosity values. The increase in the spacing between the grafted side chains allows the bonds in the system to orient better with the flow; thus, leading to a stronger shear thinning behaviour in Fig. \ref{fig:viscosity}. This is due to the local frustration of the bonds in densely-grafted bottlebrushes imposed by the existence of side chains, see Fig. S2. When the crowding around the backbone increases, the bonds become more resistant to orientation. As a result, bottlebrushes with $m>1$ exhibit higher degrees of shear thinning when $m$ is gradually increased. Star polymers and star-like bottlebrushes have fewer bonds around the frustrated region near the backbone and the average bond orientation is stronger leading to a stronger shear thinning.

\subsection{Shape Parameters}

\begin{figure}[h!]

  \includegraphics[width=\textwidth]{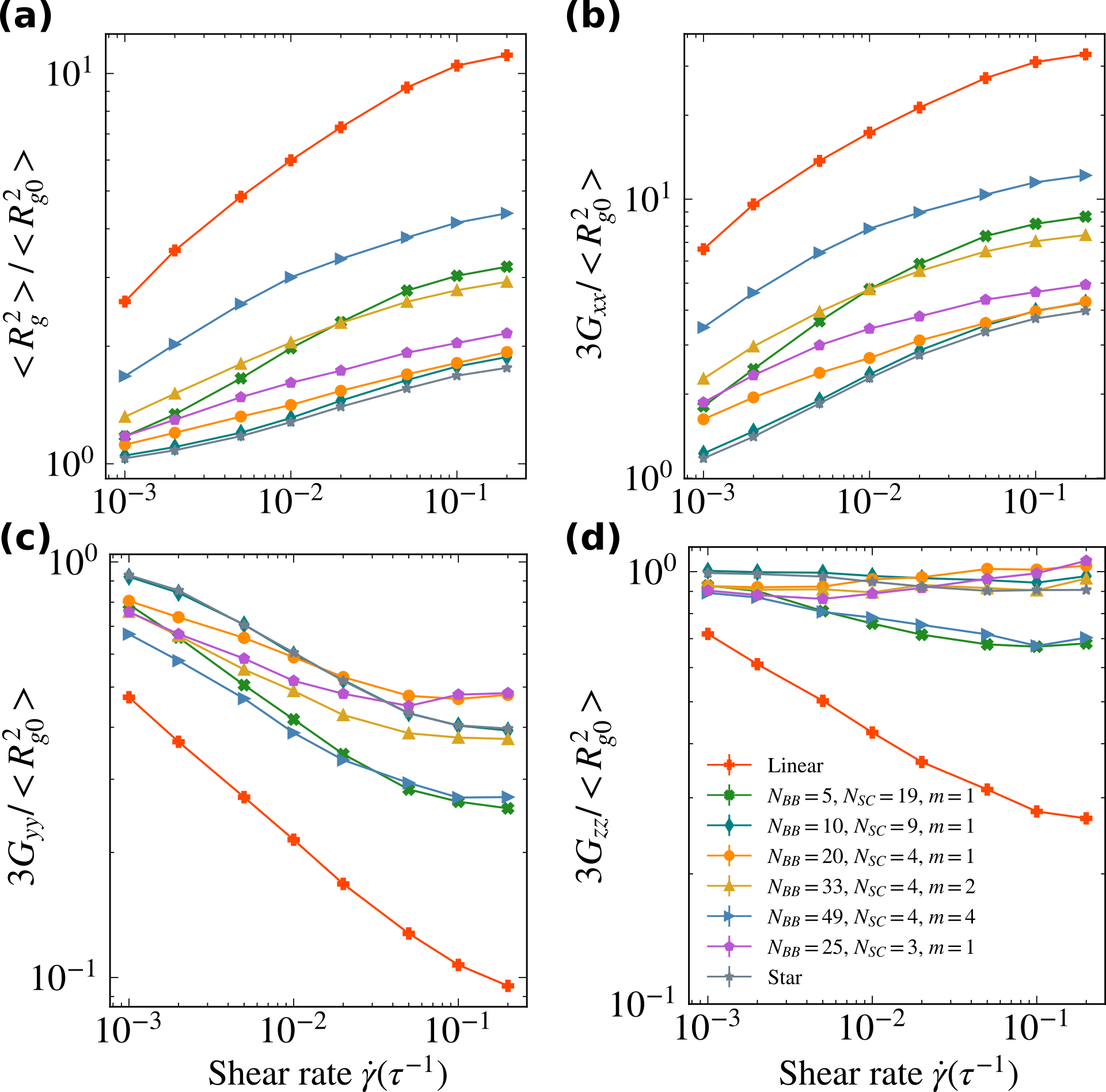}
  \caption{Radius of gyration and the diagonal components of the gyration tensor. \textbf{(a)}: Radius of gyration squared calculated as the trace of the gyration tensor. Increasing $R_g^2$ indicates an extension and stretching of polymer chains under shear. \textbf{(b)}: Increasing $x$-component of the gyration tensor represents the extension along the shear direction with increasing shear rate. \textbf{(c-d)}: Decreasing $y$ and $z$-components of the gyration tensor indicates compression with increasing shear rate on the gradient and the vorticity directions respectively.}
  \label{fig:gyration}
\end{figure}

We expect that the orientation of the molecules with the flow plays an important role in determining the system rheology, and here we calculate different shape parameters by computing the radius of gyration tensor that is commonly used to determine the shape, size, and structure of polymer chains \cite{rudnick1987shapes} . The components of the gyration tensor are given by 

\begin{equation}
    G_{i j}=\frac{1}{N} \sum_{k=1}^N\left(r_{ik}-r_i^{(C M)}\right)\left(r_{jk}-r_j^{(C M)}\right)
\end{equation}

\noindent where $N$ is the total number of beads in a single chain, $r_{ik}$ is the Cartesian components ($x, y, z$) of the $k^{th}$ monomer and $r_i^{(CM)}$ is the $i^{th}$ component of the centre of mass (CM) of the chain. We diagonalise the gyration tensor to obtain its eigenvalues $\lambda^2_{max} > \lambda^2_{mid} > \lambda^2_{min} $ in descending order. The trace of this tensor gives the radius of gyration $R_g$ for a polymer chain as $R_g^2 = \lambda^2_{max} + \lambda^2_{mid} + \lambda^2_{min}$. We show the radius of gyration $R_g^2$ and the diagonal components $G_{ii}$ in Fig. \ref{fig:gyration} which are normalised by their respective $R^2_{g0}$ values at equilibrium to identify the changes under shear with respect to equilibrium conditions. The equilibrium values of $R_{g0}^2$ are given in Table \ref{tab:equilShape} along with the average bond orientations of different architectures. Note that the curves in Fig. \ref{fig:gyration} all start from $1$ at equilibrium.

The increasing $G_{xx}$ and decreasing $G_{yy}$ (Fig. \ref{fig:gyration}) indicate that all architectures extend in the flow direction and compress in the gradient direction leading to the orientation of the whole molecule in the $\hat{x}$-direction. Note that linear chains have the highest increase in $R_g^2$ and $G_{xx}$ , while the lowest increase is observed for star polymers. Moreover, we infer a compression along the vorticity direction from the decrease in $G_{zz}$ as expected from shear-thinning polymer melts \cite{xu2016effect} . We agree with Ref. \cite{xu2016effect} that increasing branching makes $G_{zz}$ insensitive to shear. This compression is more prominent for linear chains as they have more space to compress in the $z$-direction than other architectures. The star-like (5,19,1) and the sphere-like (10,9,1) bottlebrushes also show a decrease in $G_{zz}$ at low shear rates. For the remaining systems, $G_{zz}$ either stays constant or slightly increases with the increasing shear rate. This indicates that the chains resist the orientation along the flow direction and, as a result, expand in the vorticity direction. Note that $G_{xx}$ does not correlate with the viscosity directly. Some architectures orient better in the flow direction, yet they do not show a stronger shear thinning. What we infer from the analysis of the conformational properties is that the molecules can orient with the flow as a whole; however, shape parameters based on the gyration tensor are not the key predictors of the architecture dependence on shear thinning. 

\begin{table}[h!]
\caption{\ Equilibrium properties}
  \label{tab:equilShape}
\begin{tabular}{cccc}
\toprule
Architecture            & $R_{g0}^2$ & $P_2$   \\ 
\midrule
$Linear$                 & 27.00      & -0.00081  \\    
$N_{BB}=49$, $N_{SC}=4$, $m=4$ & 16.45      & 0.00004  \\
$N_{BB}=5$, $N_{SC}=19$, $m=1$ & 14.85      & 0.00043  \\
$N_{BB}=33$, $N_{SC}=4$, $m=2$ & 12.84      & 0.00052  \\
$N_{BB}=25$, $N_{SC}=3$, $m=1$ & 10.84      & 0.00088  \\
$N_{BB}=20$, $N_{SC}=4$, $m=1$ & 9.86     & -0.00017 \\
$N_{BB}=10$, $N_{SC}=9$, $m=1$ & 9.86     & -0.00030 \\         
$Star$           & 9.58     & -0.00078 \\                  
\bottomrule
\end{tabular}
\end{table}

\subsection{Correlations of architectural parameters with viscosity}

\begin{figure}[h!]
  \includegraphics{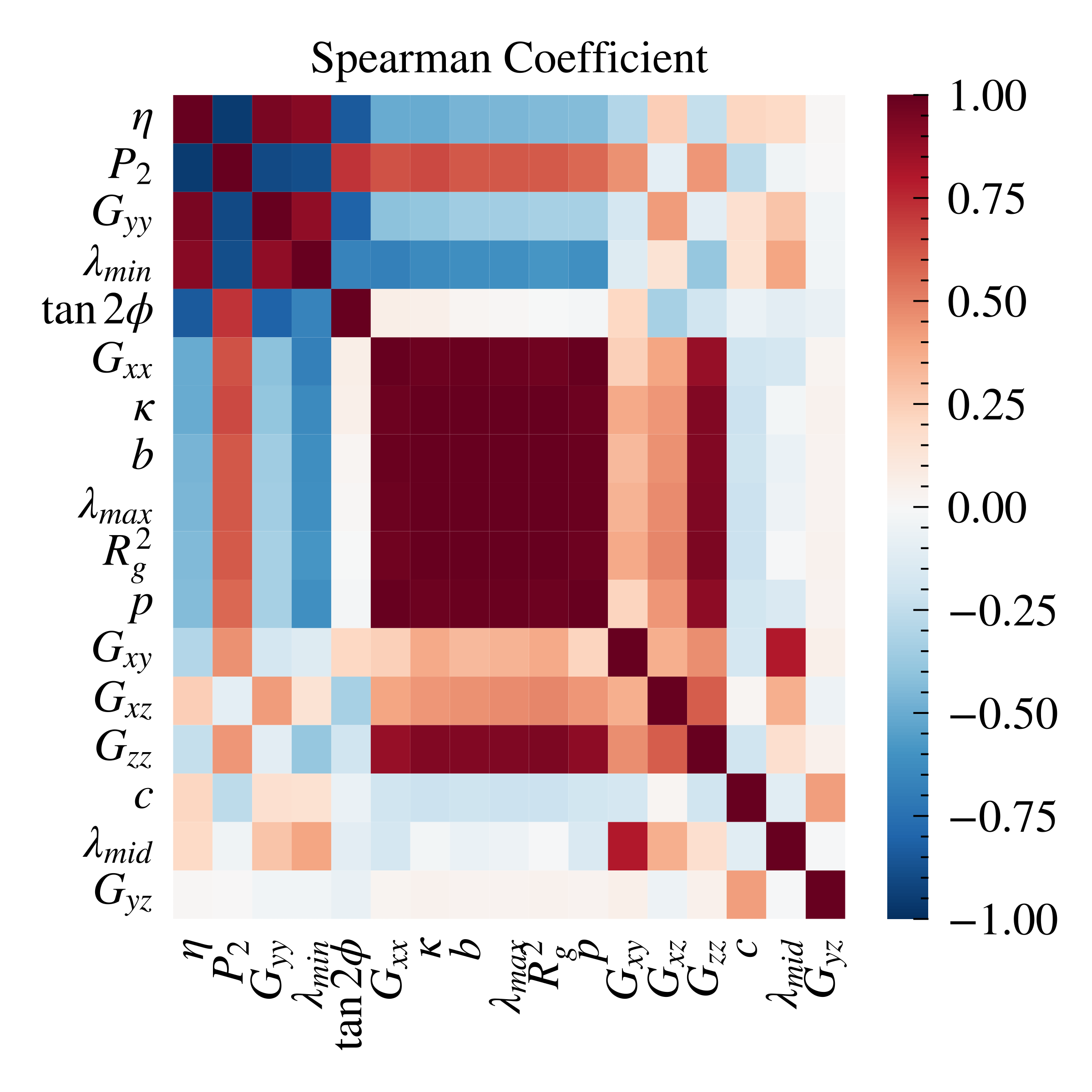}
  \caption{Spearman correlation coefficient between different parameters. We report the correlations between shape parameters and the viscosity that are given in Table \ref{tab:correlations} sorted by decreasing absolute value. We also see the correlations between different shape parameters. The red block in the middle of the map corresponds to high correlations of shape parameters expressed in terms of gyration tensor eigenvalues. Hence, we expect them to be somewhat correlated. Smaller values of the relative shape anisotropy ($\kappa$) determine how isotropic a configuration is. Similarly for asphericity ($b$) and acylindricity ($c$). Detailed definitions of these parameters are given in the Methods.}
  \label{fig:spearman}
\end{figure}

One could think that it is not $G_{ii}$, but some geometrical shape combination of them, e.g. asphericity, which correlates with the shear thinning. We test various architectural parameters which are a function of gyration tensor components with different weights and combinations. Indeed, all shape parameters investigated behave as expected under flow. For instance, the asphericity increases and this increase is more prominent for stars and star-like systems than bottlebrush-like systems. However, no other shape parameter correlates as well with shear thinning as $P_2$, across different architectures. We quantify this by the Spearman coefficient given in Fig. \ref{fig:spearman}, and the full data sets of these parameters as a function of shear rate for all architectures are reported in Fig. S3-S4. We consider these parameters as a function of shear rate, which is an independent variable, and compute the correlation between two data sets by sorting them from low to high shear rate. The correlations between these parameters and the viscosity are reported in Table \ref{tab:correlations} in descending order of the absolute value. Although $G_{yy}$ correlates with viscosity as much as $P_2$ does, it cannot explain the ranking of viscosities for different architectures at high shear rates. On the other hand, $P_2$ is able to describe this trend for all architectures. A natural parameter for this study would be $G_{xx}$ as we apply the shear in the $x$-direction; however, it does not correlate with viscosity. Hence, the reason for high correlations in $G_{yy}$ is a secondary effect as the chains have to compress in the gradient direction when the shear rate is increased. There is no strong correlation between the viscosity and all the other quantities reported in Table \ref{tab:correlations}.

Our findings align with recent experimental results and provide a precise picture of the molecular rearrangements giving rise to the observed rheological behaviour of our melts.  Like Zografos \textit{et al.} \cite{zografos2023star} and Lopez \textit{et al.} \cite{lopez2019extensional} , for densely grafted bottlebrushes we identify a star-like to bottlebrush-like transition with power law exponents for the flow curves progressively decreasing and  converging to a Zimm scaling of $\sim -0.33$ with increasing $N_{BB}$ \cite{zimm1956dynamics} , even if in our case the total molecular weight is constant instead of progressively increasing $N_{BB}$ at fixed side chain length. We attribute the shear thinning behaviour to the orientation of the individual bonds rather than the backbone relaxation, in line with the computational findings of \textit{Qu et al.} \cite{qu2022dispersion} . This idea would also be in line with the recent findings that a monomeric friction coefficient in Brownian dynamics simulations of Fraenkel chains is necessary to correctly capture experimental viscosity data for unentangled linear polymers under flow \cite{ianniruberto2019modeling, ianniruberto2020origin, matsumiya2018nonlinear} . Yet, this might be caused by the high shear rates accessible to our simulations, close to the inverse of the bond relaxation times $\tau_b$, while different architectural effects might play a more important role at lower shear rates that we could not access. Nevertheless, we agree with the idea that for densely grafted bottlebrushes the side chains are not able to align well in the flow direction, which might lead to strain-hardening in extensional flow \cite{zografos2023star,lopez2019extensional} . Still, we attribute this to the steric repulsion of side chains close to the backbones, rather than to interdigitation. In fact, both in densely and loosely grafted bottlebrushes the side chains still are somewhat aligned with the flow ($P_2>0$, see Fig. S2) and not really interlocked in the direction perpendicular to the flow.
\begin{table}[h]
\caption{\ Spearman correlation coefficient between different parameters and viscosity}
  \label{tab:correlations}
\begin{tabular}{cc}
\toprule
Parameter        & Spearman Coefficient \\
\midrule
$P_2$            & -0.96                \\
$G_{yy}$         & 0.95                 \\
$\lambda_{min}$  & 0.89                 \\
$\tan 2\phi$     & -0.81                \\
$G_{xx}$         & -0.51                \\
Shape Anisotropy ($\kappa$) & -0.50                \\
Asphericity ($b$)      & -0.48                \\
$\lambda_{max}$  & -0.47                \\
$R_g^2$          & -0.46                \\
Prolateness ($p$)      & -0.43                \\
$G_{xy}$         & -0.30                \\
$G_{xz}$         & 0.23                 \\
$G_{zz}$         & 0.23                 \\
Acylindricity ($c$)    & -0.23                \\
$\lambda_{mid}$  & 0.18                 \\
$G_{yz}$         & 0.05                 \\
\bottomrule
\end{tabular}
\end{table}

\FloatBarrier

\section{Conclusions}

We have studied the shear thinning of bottlebrush polymer melts under shear flow using non-equilibrium molecular dynamics simulations. We explore a wide range of architectures at fixed molecular weight, transitioning from linear chains to combs, to densely grafted bottlebrushes, to star-like and star polymers.  We report a range of shear thinning exponents from a Rouse regime for linear and star-like systems to a Zimm regime for cylindrical bottlebrushes, observing a star-like to bottlebrush-like transition in line with recent experiments\cite{zografos2023star} . We observe directly the alignment of all molecules with the flow direction at an increasing shear rate, which is more pronounced for linear chains and for bottlebrush-like polymers than for stars due to their anisotropic shape. We quantify molecular alignment in the form of several parameters extracted from the gyration tensor such as the principal components along and perpendicular to the flow, and shape parameters such as asphericity or prolateness. Yet, in the range of shear rates considered, the alignment of molecular bonds explains the architecture effect on shear thinning, rather than the molecular alignment. Linear chains show the strongest shear thinning because all bonds can easily align with the flow, while this is hindered for the side chains of cylinder-like bottlebrushes, which then shear thin the least. Lopez \textit{et al.} \cite{lopez2019extensional} and Zografos \textit{et al.} \cite{zografos2023star} attribute the strain hardening observed in extensional flow in the cylindrical bottlebrush regime to this bottlebrush backbone alignment and increased friction due to side chain interdigitation; we see a similar alignment, but our results suggest that the effect on rheology is rather caused by the orientation of the backbone and side chain bonds. Overall, our findings line up with the intuition obtained from experimental results on the rheology of bottlebrushes in different regimes but offer important insight into the molecular configurations of different architectures under flow that is crucial to determine and control the flow properties of bottlebrush melts by tuning their architecture. It will be important in future work to extend these findings to a wider range of molecular weights and shear rates, and to eventually quantify analytically the architecture-structure-property relationships under flow for this class of branched polymers.


\section{Code availability}

In an effort to promote Open Science practices, the LAMMPS code used to perform our simulations is available on the GitHub page of our group \url{https://github.com/giuntoli-group/bottlebrush-shear-thinning}

\begin{acknowledgement}

We thank the Center for Information Technology of the University of Groningen for their support and for providing access to the Hábrók high performance computing cluster. This work made use of the Dutch national e-infrastructure with the support of the SURF Cooperative using grant no. EINF-2648. We thank Daniele Parisi for the useful discussions and Ya\u{g}mur \"{O}zt\"{u}rk for proofreading.

\end{acknowledgement}

\begin{suppinfo}

This material is available for download from \url{http://pubs.acs.org}. 

\end{suppinfo}

\bibliography{references}

\end{document}


\newpage

\subsection{Characteristic Relaxation Time Scales}

\begin{figure}[]
  \includegraphics[scale=1.25]{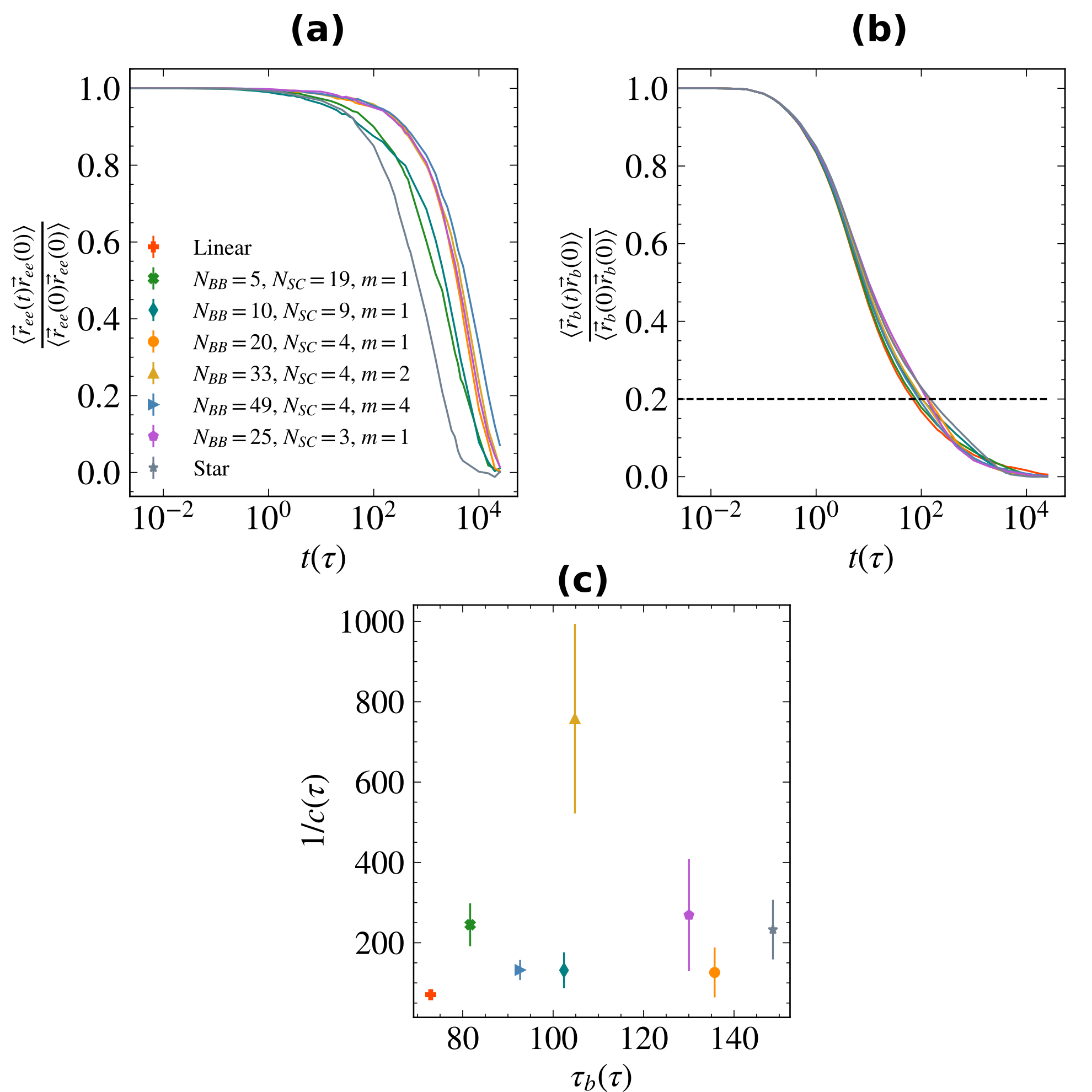}
  \caption{\textbf{(a)} Auto-correlation function of the longest chains in a molecule. For long bottlebrushes, $\vec{r}_{ee}$ refers to the end-to-end position vector of the backbone or the side chain depending on the number of beads on each sub-chain. For stars, it is the position vector from the core bead to the terminal bead in one arm. For linear chains, it is the end-to-end position vector of the whole chain. The end result is obtained by averaging  $\vec{r}_{ee}$ for all molecules in the system. \textbf{(b)} Auto-correlation function of the bond vectors $\vec{r}_{b}$. We compute the correlations by considering all bonds in a given system. \textbf{(c)} The relation between the bond relaxation time $\tau_b$ and the fit parameter $c$. $\tau_b$ is the time at which the bond auto-correlation function reaches the value $0.2$. Although there is no clear correlation between the two, our fit parameter is of the same order of magnitude as the bond relaxation time. }
  \label{suppfig:relaxation_times}
\end{figure}

We estimate the relaxation times of different architectures by computing the end-to-end auto-correlation function in Fig. \ref{suppfig:relaxation_times}. For bottlebrushes, estimating this time is complicated since there are different relaxation times related to different parts of the chain. Hence, we consider the longer sub-chain in determining this relaxation time. We compute $\vec{r}_{ee}$ of all sub-chains that satisfy $max(N_{BB}, N_{SC})$. After obtaining the auto-correlation function, the relaxation time is estimated as the time at which this function decays to $0.2$. 

\FloatBarrier

\subsection{Bond orientation of backbones and side chains}

\begin{figure}[]
  \includegraphics[scale=1.25]{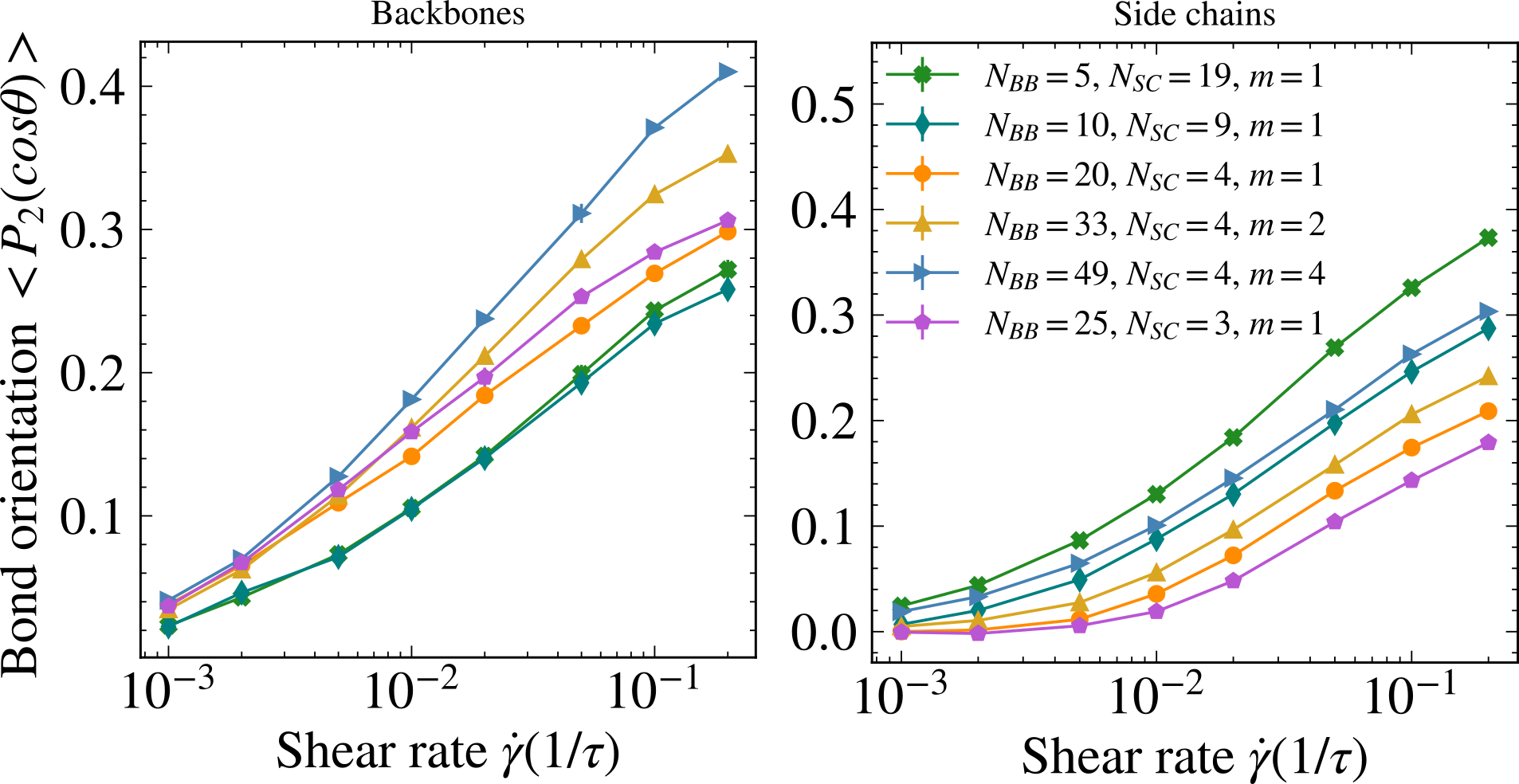}
  \caption{Bond orientation parameter $P_2$ of backbones (left) and side chains (right) are calculated separately. Backbone and side chain bonds orient more easily as the number of grafted side chains is decreased at fixed number of side chain beads $N_{SC}$.}
  \label{suppfig:p2_bb_sc}
\end{figure}

We compute the $P_2$ parameter separately for bottlebrush backbones and side chains in Fig. \ref{suppfig:p2_bb_sc}, excluding the linear chain and the star polymer that don't have both types of bonds. We obseve that both the backbone bonds and side chains bonds orient more easily as we start to decrease the grafting density of side chains without changing their length $N_{SC}$. 

\FloatBarrier
\subsection{Architectural Parameters}

\begin{figure}[]
  \includegraphics[scale=1.75]{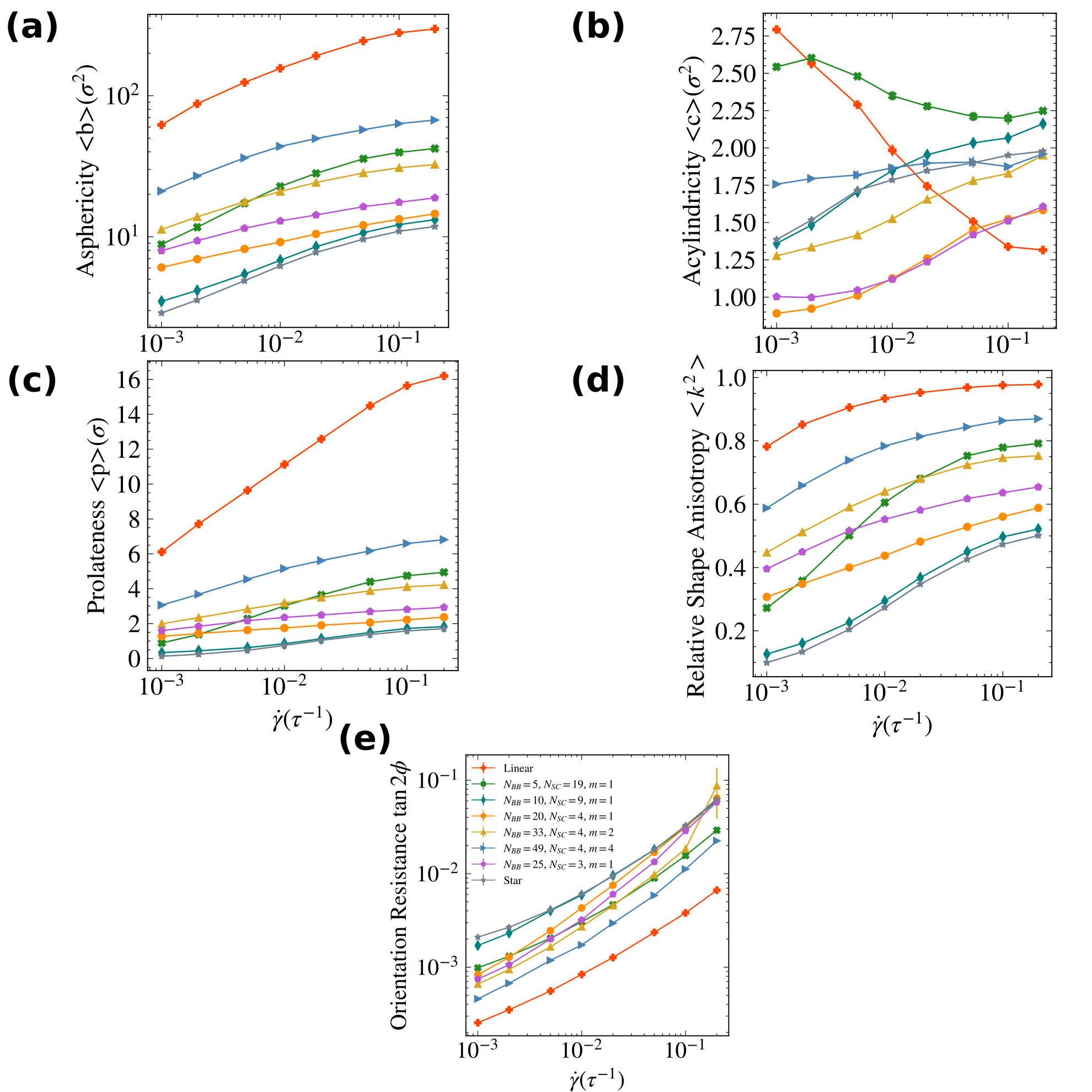}
  \caption{\textbf{(a)} \textbf{Asphericity} increases with increasing shear rate. The chains are more isotropic at low shear rates and they are deformed differently depending on the architecture.\textbf{(b)} \textbf{Acylindricity}. Bottlebushes are more cylindrical under equilibrium conditions and their shape is deformed under shear, whereas linear chains are more globular at equilibrium. They become more cylindrical as the chains are extended under high shear deformation \textbf{(c)} \textbf{Prolateness}. Larger values indicate more prolate, and smaller ones indicate more oblate shapes. \textbf{(d)} \textbf{Relative shape anisotropy}. All architectures are more isotropic at equilibrium which then is reduced with increasing shear rate. \textbf{(e)} \textbf{Orientation resistance parameter}   }
  \label{suppfig:shape_params}
\end{figure}

We compute several architectural parameters based on the gyration tensor in Fig. \ref{suppfig:shape_params}. These parameters are a function of the gyration tensor eigenvalues.

\FloatBarrier

\begin{figure}[]
  \includegraphics[scale=1.5]{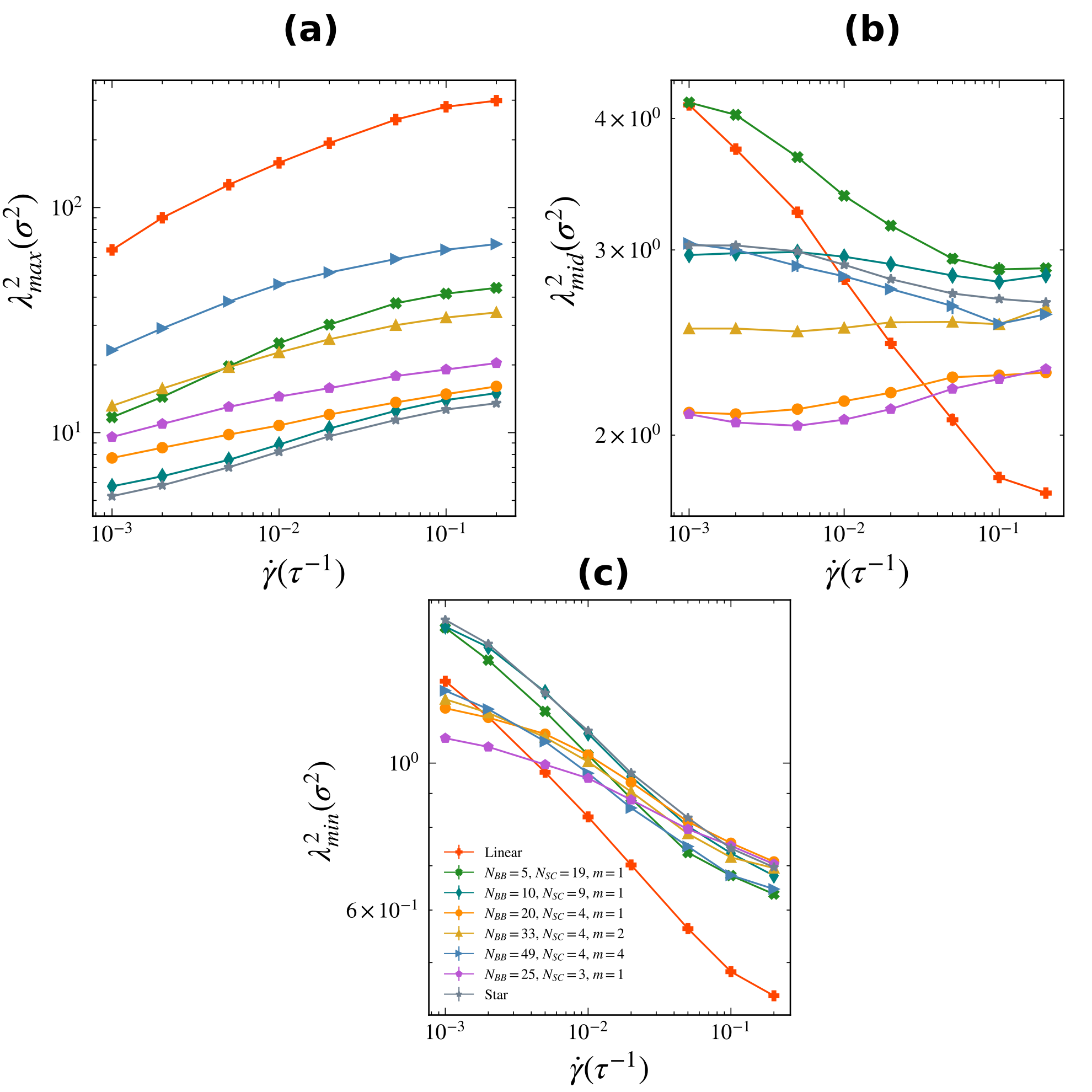}
  \caption{\textbf{(a)} \textbf{Largest eigenvalue} is directly proportional to the size of the polymer. It also defines the principal symmetry axis, i.e. the corresponding eigenvector, of the molecule. $\lambda^2_{max}$ increases with shear rate for all architectures indicating an extension along the molecule's symmetry axis. Significantly larger  $\lambda^2_{max}$ compared to other eigenvalues indicates an elongated shape.  \textbf{(b)} \textbf{Middle eigenvalue}   \textbf{(c)} \textbf{Smallest eigenvalue} }
  \label{suppfig:eigenvalues}
\end{figure}

Eigenvalues of the gyration tensor (see Fig. \ref{suppfig:eigenvalues}) help us identify the molecular size, shape, and symmetry. They also give insights on the molecular conformation.  

\FloatBarrier